\documentclass[12pt]{article}
\newcommand{\be}{\begin{equation}}
\newcommand{\ee}{\end{equation}}
\newcommand{\G}{\Gamma}
\newcommand{\g}{\gamma}
\begin{document}
\thispagestyle{empty}
\begin{flushright}
January 2002\\
\end{flushright}
\vspace{1.5cm}
\begin{center}
{\Large\bf Dynamical masses of quarks in \\[0.2truecm]

quantum chromodynamics }\\[1.7truecm]
{\large \bf N.V. Krasnikov and A.A. Pivovarov
}
\vspace{0.7cm}

Institute for Nuclear Research of the\\
Russian Academy of Sciences, Moscow 117312, Russia
\end{center}

\vspace{1truecm}
\begin{abstract}
\noindent 
Using Dyson-Schwinger equations we obtain
an ultraviolet asymptotics for the dynamical mass of quark in QCD
\[
m_D(p)=\frac{\langle \bar q q \rangle }{3}
\frac{\bar g^2(p^2/\mu^2,g^2_R)}{p^2}
\left(\frac{\bar g^2(p^2/\mu^2,g^2_R)}{g^2_R}\right)^{-d}\, .
\]
We also determine a numerical value for the $\pi$ meson decay constant
$f_\pi$.
\end{abstract}

\vskip 5.3cm
\centerline{Electronic version of the paper 
published in Russ.~Phys.~J.  25, 55 (1982).}

\newpage
As is well known, strong interactions at high energies are
approximately invariant with respect to the chiral symmetry group
$SU_L(N)\otimes SU_R(N)$ ($N$ is the number of different quark
species).  The Lagrangian of QCD -- the most popular field-theoretical
model of strong interactions -- has the chiral symmetry group
$SU_L(N)\otimes SU_R(N)$ in approximation of massless quarks.  The
success of current algebra and PCAC hypothesis shows that the chiral
symmetry  $SU_L(N)\otimes SU_R(N)$ is spontaneously broken down to
$SU(N)$ symmetry. The spontaneous violation of the chiral symmetry
reveals itself in nonvanishing vacuum expectation values  $\langle
\bar q q \rangle$ and also in the appearance of dynamical mass for
originally massless quarks.

The problem of dynamical mass generation in massless theories has a
long history. First, in refs.~\cite{arb,nambu}        it has been
shown in some concrete models that the generation of the dynamical
mass can take place in quantum field theory. However, the consistent
study of the problem of dynamical mass generation requires going
beyond perturbation theory. It is obvious that this problem is rather
nontrivial in the field theory because of the absence of the
consistent methods different from perturbation theory.
         
The purpose of the present paper is to investigate a possibility of
generation of the dynamical mass of quarks in QCD. Using the
Dyson-Schwinger equations we show that  one can reliably determine the
dependence of the dynamical mass of quarks $m(p)$ on the momentum $p$
in the ultraviolet domain due to the property of asymptotic freedom of
strong interactions.  In the following we will work in Euclidean
space-time and in the transverse gauge (Landau gauge). The use of this
gauge is convenient since the propagator of the massless quark is not
renormalized in the leading logarithmic approximation. 

The Dyson-Schwinger equation for the quark propagator $G(p)$ has the
form 
\be
\label{dseqn}
G^{-1}(p) = S^{-1}(p)+\frac{g^2}{(2\pi)^4}
\int d^4k \, \G_\mu^a(p,k) G(k)
D_{\mu\nu}(p-k) \g_\nu (\lambda^a/2) .
\ee
Here $S(p)$ is the propagator of free quarks, $\lambda^a/2$ are the 
generators
of the fundamental representation of the group $SU(N)$. 
Equation~(\ref{dseqn})
is written for bare quantities. After performing the renormalization
we find 
\[
\label{dseqnREN}
Z_2^{-1}G^{-1}_R(p) = S^{-1}(p)+\frac{g_R^2}{(2\pi)^4}
\int d^4k \, \G_\mu^{aR}(p,k) G_R(k)
D_{\mu\nu}^R(p-k) \g_\nu (\lambda^a/2) ,
\]
where $Z_2$ is the renormalization constant of the quark
propagator. At the lowest order in $g^2_R$ in the Landau gauge the
renormalization constant of the quark propagator is equal to one
$Z_2=1+O(g^4_R)$. Therefore in the leading logarithmic approximation 
we can set $G^{-1}(p)=m(p)-\hat p$. For the quantity $m(p)$ 
we obtain the equation of the form 
\[
m(p^2)=\frac{4}{(2\pi)^4}\int d^4k \frac{m(k^2) 
{\bar g}^2((p-k)^2)}{(m^2(k^2)+k^2)(p-k)^2}
\]
where 
\[
\bar g^2(p^2)=\bar g^2(p^2/\mu^2,g^2_R)=g^2_R/\left(1+(11-2N/3)
\frac{g^2_R}{16\pi^2}\ln\frac{p^2}{\mu^2}\right) 
\]
is the effective quark-gluon coupling constant. We split the region of
integration into two sub-domains determined by the requirements 
$k^2>p^2$ (sub-domain~(I))
and $k^2<p^2$ (sub-domain~(II)) 
\[
\int d^4k =\int_{k^2>p^2,~(I)} d^4k +\int_{k^2<p^2,~(II)} d^4k \, .
\]
Then, taking into account that 
\[
x\frac{d}{dx}\bar g^2(x)=-\frac{33-2N}{48\pi^2}(\bar g^2(x))^2
\]
we obtain the equality $\bar g^2[(p-k)^2]=\bar g^2(k^2)$
up to the terms of the next order in 
$g^2$ in the sub-domain (I) where $k^2>p^2$ and the equality
$\bar g^2[(p-k)^2]=\bar g^2(p^2)$
with the same accuracy in the sub-domain (II) where $p^2>k^2$. 

Setting $p^2=x$ we obtain after integrating over the angular variables
\be
\label{eq2}
m(x)=\frac{1}{4\pi^2}\int_x^\infty 
\frac{\bar g^2(z)m(z)dz }{m^2(z)+z}
+\frac{1}{4\pi^2}\frac{\bar g^2(x)}{x}
\int_0^x \frac{m(z)z dz }{m^2(z)+z}\, .
\ee
After some transformations of eq.~(\ref{eq2})
one finds  
\[
\frac{d}{dx}\left(\frac{4\pi^2 x^2 (dm/dx)}{\bar g^2(x)}\right)
=-\frac{m(x) x}{m^2(x)+x}\, . 
\]
Taking into account that 
\[
(33-2N)\bar g^2(x)=48\pi^2/\ln(x/\Lambda^2)
\]
and changing the variable
$x/\Lambda^2=\exp(t)$ we obtain
\be
\label{sigmaeqn}
\frac{d^2\Sigma(t)}{d t^2}+(1+1/t)\frac{d\Sigma(t)}{d t}
+d\frac{\Sigma(t)}{t}(1+\Sigma^2(t)\Lambda^{-2}\exp(-t))^{-1}
=0
\ee
where $d=12/(33-2N)$, $m(x)\equiv\Sigma(t)$ with $t=\ln (x/\Lambda^2)$.

We are looking for a solution of eq.~(\ref{sigmaeqn})
which is bounded at $t\to \infty$, therefore in the asymptotic regime
one can neglect the nonlinearity of the equation. The linear equation
has the solution 
\[
\Sigma(t)=\exp(-t/2-\frac{1}{2}\ln t)
\left(\tilde C_1\, W_{-d+1/2,0}(-t)+\tilde C_2\, W_{d-1/2,0}(t)\right)
\]
where 
$W_{\lambda,\mu}(z)$ are Whittaker functions with the asymptotic 
behavior for large $z$
\be
\label{asympW}
W_{\lambda,\mu}(z)|_{|z|\to \infty}=\exp(-z/2) z^\lambda[1+O(1/z)]\, .
\ee
Using eq.~(\ref{asympW}) 
we obtain the asymptotics of two linearly independent solutions of
eq.~(\ref{sigmaeqn}):
\[
m_1(x)|_{x\to \infty} \sim C_1 \ln^{-d}(x/\Lambda^2)
\]
and
\[
m_2(x)|_{x\to \infty} \sim (C_2/x) \ln^{d-1}(x/\Lambda^2)\, .
\]

For the determination of the constants $C_1$ and $C_2$ we use
the comparison with perturbation theory. We require smallness of the
nonperturbative
corrections in the asymptotic domain in the form 
\[
m(p^2)-\bar m(p^2)=\bar o(\bar m(p^2)), \quad p^2\to \infty
\]
where $\bar m(p^2)=m_R\left(\bar g^2(p^2)/\bar g^2(\mu^2)\right)^d$
is the usual expression for the effective quark mass in the
renormalization group analysis.
Then 
\[
 C_1=m_R \ln^{d}(\mu^2/\Lambda^2)
\]
 and 
\be
\label{massoperator}
m(p^2)=m_R\left(\bar g^2(p^2)/\bar g^2(\mu^2)\right)^d
-\frac{C}{p^2}\bar g^2(p^2)
\left(\bar g^2(p^2)/\bar g^2(\mu^2)\right)^{-d}\, .
\ee
The second term in this expression describes the ultraviolet 
asymptotics of the dynamical mass of the light quark.
For the determination of the constant $C$ we use the perturbation
theory with phenomenological account for the nonperturbative effects
through $\langle \bar q q\rangle \ne 0$.

In the second order of the perturbation theory with the nonvanishing
vacuum expectation value of the bilinear quark operator 
$\langle \bar q q\rangle$ one has 
\[
{\rm tr}~G(p) =-\frac{4\langle \bar q q\rangle}{3}g_R^2/(p^2)^2\, ,
\]
on the other hand the use of the explicit form of the mass 
operator from eq.~(\ref{massoperator}) gives 
\[
{\rm tr}~G(p) =-4C g_R^2 /(p^2)^2\, .
\]
Comparing these two expressions we obtain 
\be
\label{constC}
C=\langle \bar q q\rangle/3 \,  .
\ee

With the expression for the dynamical mass given in
eq.~(\ref{massoperator}) 
where the constant $C$ is fixed by the relation from
eq.~(\ref{constC}) 
we compute the pion 
decay constant $f_\pi$ defined through 
\[
\langle 0|A^a_\mu(0)|\pi^b(\vec{k})\rangle
=i k_\mu f_\pi \delta^{ab}, 
\quad  k^2=m_\pi^2,
\]
where $|\pi^a(\vec{k})\rangle$ is the one-particle pion state, $A^a_\mu(0)$
is the isotopic vector axial current. The representation for the pion 
decay constant $f_\pi$ was found in ref.~\cite{jj}
and used earlier for a similar calculation in ref.~\cite{pagels}
\be
\label{fpi}
f_\pi^2=\frac{3}{(2\pi)^2}\int_0^\infty dx 
\frac{x\left(m(x)-\frac{x}{2}\frac{d m(x)}{dx}\right)m(x)}
{(x+m^2(x))^2}\, .
\ee

For determining the numerical values one has to know the magnitude of
the vacuum expectation value $\langle \bar q q\rangle$. 
The methods of
current algebra allow one to relate  $\langle \bar q q\rangle$ 
to the current masses of light quarks that leads
to the estimate $\langle \bar q q\rangle=-(250~{\rm MeV})^3$.
Then the relation from eq.~(\ref{fpi}) gives $f_\pi=77~{\rm MeV}$. 
Using the experimental value $f_\pi=93~{\rm MeV}$ one can turn the
problem around and determine 
$|\langle \bar q q\rangle|=(310~{\rm MeV})^3$. 
Note that in this way one cannot determine the sign of the
vacuum expectation value $\langle \bar q q\rangle$.

\end{document}